\documentclass[]{spie}  

\usepackage{amsmath,amsfonts,amssymb}
\usepackage{graphicx}
\usepackage[colorlinks=true, allcolors=blue]{hyperref}
\usepackage{float}
\usepackage{pifont}

\title{Laboratory test of the VIS detector system of SOXS for the ESO-NTT telescope}

\author[a,b]{Rosario Cosentino}
\author[a]{Marcos Hernandez}
\author[a]{Hector Ventura}
\author[c]{Sergio Campana}
\author[d]{Riccardo Claudi}		
\author[e]{Pietro Schipani}	
\author[c]{Matteo Aliverti}		
\author[d]{Andrea Baruffolo}	
\author[f,g]{Sagi Ben-Ami}		
\author[h]{Federico Biondi}		
\author[e]{Giulio Capasso}		
\author[i]{Francesco D'Alessio}
\author[c]{Paolo D'Avanzo}		
\author[f]{Ofir	Hershko}		
\author[j,k]{Hanindyo Kuncarayakti}
\author[c]{Marco Landoni}			
\author[b]{Matteo Munari}			
\author[l,m]{Giuliano Pignata}		
\author[n]{Adam Rubin}				
\author[b,o]{Salvatore Scuderi}		
\author[i]{Fabrizio Vitali}			
\author[p]{David Young}				
\author[q]{Jani	Achrén}				
\author[l,r]{José Antonio Araiza-Duran}
\author[s]{Iair	Arcavi}			
\author[m,t]{Anna Brucalassi}	
\author[f]{Rachel Bruch}		
\author[d]{Enrico Cappellaro}	
\author[e]{Mirko Colapietro}	
\author[e]{Massimo Della Valle}	
\author[d]{Marco De Pascale}	
\author[b]{Rosario Di Benedetto}
\author[e]{Sergio D'Orsi}		
\author[g]{Avishay Gal-Yam}		
\author[c]{Matteo Genoni}		
\author[j,k]{Jari Kotilainen}	
\author[u]{Gianluca Li Causi}
\author[e]{Laurent Marty}	
\author[j]{Seppo Mattila}		
\author[g]{Michael Rappaport}	
\author[d]{Kalyan Radhakrishnan}
\author[d]{Davide Ricci}		
\author[c]{Marco Riva}			
\author[d]{Bernardo Salasnich}
\author[d]{Alessandra Slemer}	
\author[p]{Stephen Smartt}
\author[b]{Ricardo Zanmar Sanchez}
\author[v]{Maximilian Stritzinger}
\affil[a]{INAF - Fundaci\'{o}n Galileo Galilei, Bre\~{n}a Baja, Spain}
\affil[b]{INAF - Osservatorio Astrofisico di Catania, Catania, Italy}
\affil[c]{INAF - Osservatorio Astronomico di Brera, Merate, Italy}
\affil[d]{INAF - Osservatorio Astronomico di Padova, Padua, Italy}
\affil[e]{INAF - Osservatorio Astronomico di Capodimonte, Naples, Italy}
\affil[f]{Weizmann Institute of Science, Rehovot, Israel} 
\affil[g]{Harvard-Smithsonian Center for Astrophysics, Cambridge, USA}
\affil[h]{Max-Planck-Institut für Extraterrestrische Physik, Garching, Germany}
\affil[i]{INAF - Osservatorio Astronomico di Roma, Rome, Italy}
\affil[j]{Tuorla Observatory, Department of Physics and Astronomy, University of Turku, Turku, Finland}
\affil[k]{FINCA - Finnish Centre for Astronomy with ESO, Turku, Finland}
\affil[l]{Millennium Institute of Astrophysics (MAS), Santiago, Chile}
\affil[m]{Universidad Andres Bello, Santiago, Chile}
\affil[n]{European Southern Observatory, Garching, Germany}
\affil[o]{INAF - Istituto di Astrofisica Spaziale e Fisica Cosmica, Milano, Italy}
\affil[p]{Queen's University Belfast, Belfast, UK}
\affil[q]{Incident Angle Oy, Turku, Finland}
\affil[r]{Centro de Investigaciones en Optica A. C., León, Mexico}
\affil[s]{Tel Aviv University, Tel Aviv, Israel}
\affil[t]{INAF-Osservatorio Astrofisico Arcetri, Firenze, Italy}
\affil[u]{INAF - Istituto di Astrofisica e Planetologia Spaziali, Rome , Italy}
\affil[v]{Aarhus University, Aarhus, Denmark}
\authorinfo{Further author information: (E-mail: cosentino@tng.iac.es, Telephone: +34 922433642\\ }
 
\begin{document} 
\maketitle

\clearpage
\begin{abstract}
SOXS is the new spectrograph for the ESO NTT telescope able to cover the optical and NIR bands thanks to two different arms: the UV-VIS (350-850 nm), and the NIR (800-2000 nm). In this article, we describe the final design of the visible camera cryostats, the test facilities for the CCD characterization, and the first results with the scientific detector.
The UV-VIS detector system is based on a e2v CCD 44-82, a custom detector head coupled with the ESO Continuous Flowing Cryostat (CFC) cooling system and the New General Detector Controller (NGC) developed by ESO.
The laboratory facility is based on an optical bench equipped with a Xenon lamp, filter wheels to select the wavelength, an integrating sphere, and a calibrated diode to measure the flux.
This paper outlines the visible camera cryostat, the test facilities for the CCD characterization and the first results with the scientific detector in the laboratory and after the integration to the instrument.
\end{abstract}

(Ref. ~\citenum{soxsold,soxscosentino,soxsaliverti,soxscapasso,soxssanchez,soxsschipani,ricci2020,soxsrubin,ngcpaper,aliverti,claudi,cosentino2020,scuderi2022,shipani2022,rubin2022,claudi2022,aliverti2022}).




\section{The UV-VIS detector system}
\begin{figure} [H]
   \begin{center}
   \begin{tabular}{c} 
   \includegraphics[height=11 cm]{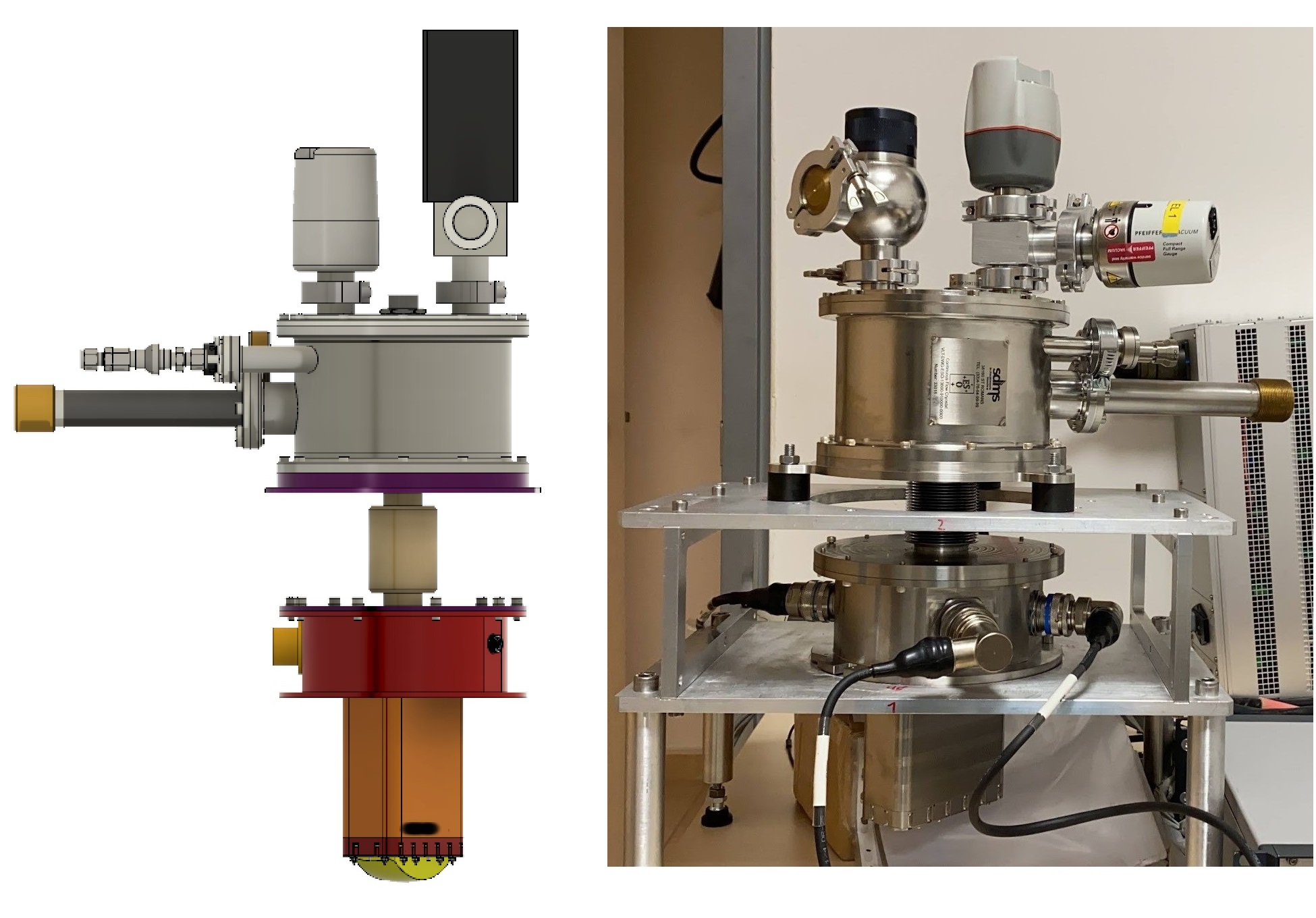}
   \end{tabular}
   \end{center}
   \caption[VIS Camera] 
   { \label{fig1} The design of the chamber and the chamber in the test bench.}
   \end{figure}

The detector head is a custom design, coupled with an ESO cooling system, based on Continuous Flow Cryostats (CFC) and a Programmable Logic Controller (PLC) for the temperature measurement and control (Figure ~\ref{fig1}).

The e2v detector is a back illuminated CCD with a 15 $\mu$m  square pixel and an image area of 30.7 x 61.4 mm. 
The high quantum efficiency (QE) in the spectral response of the spectrograph (350-850 nm) and their geometric characteristics (pixel size and dimensions) make this CCD the most suitable for our instrument.

\begin{figure} [H]
   \begin{center}
   \begin{tabular}{lcc} 
   \includegraphics[height=7 cm]{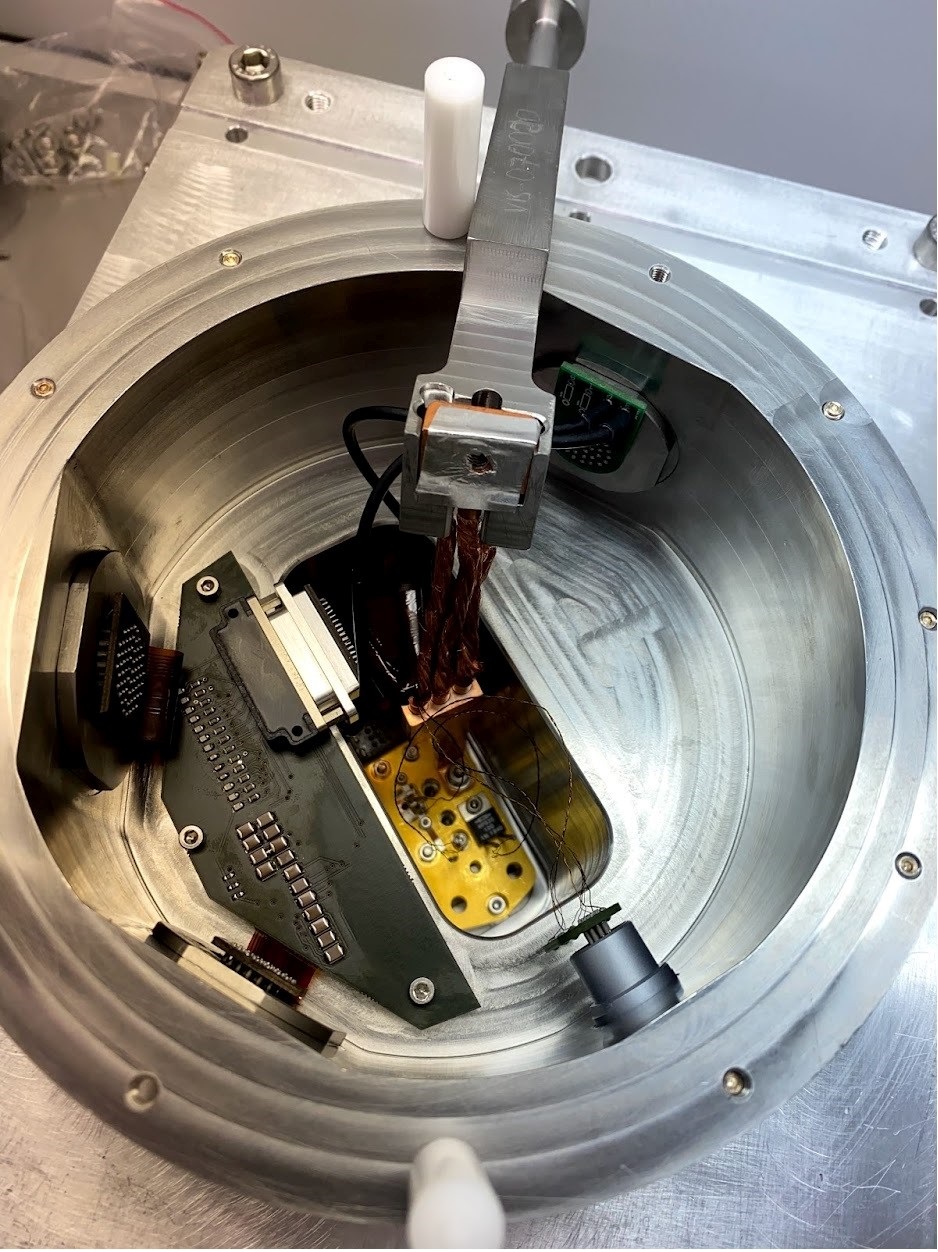}
   \includegraphics[height=7 cm]{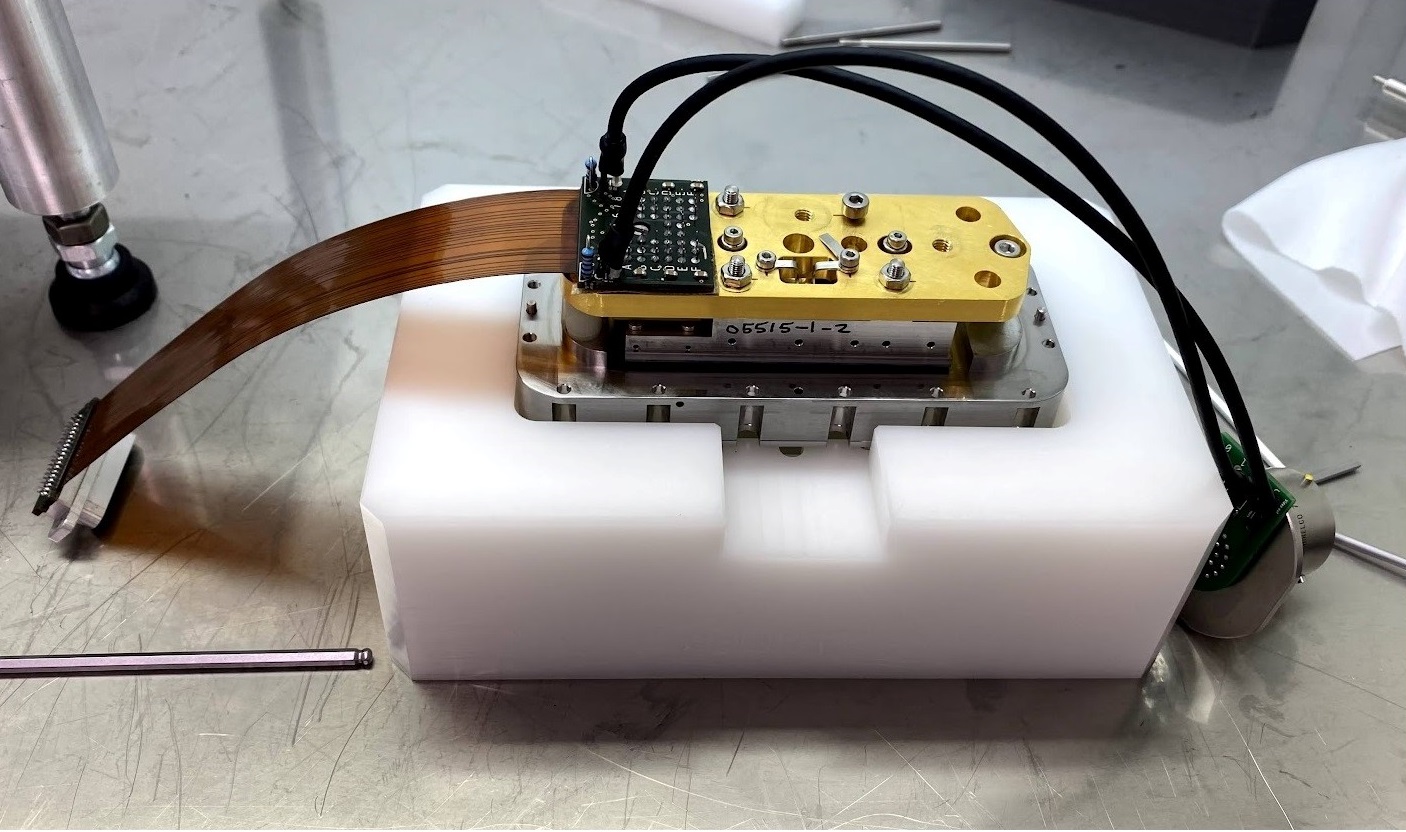}
   \end{tabular}
   \end{center}
   \caption[VIS Camera] 
   { \label{fig2} The CCD assembled in the VIS chamber (left) and The CCD cold finger with the flat PCB (right). }
   \end{figure}

The detector selected for the UV-VIS arm is an e2v CCD44-82. This detector is a high performance, back illuminated CCD with a 15.0 µm square pixel and an image area of 30.7x61.4 mm and is characterized for the high Quantum Efficiency (QE) for the instrument requirements. The detector wiring is based on a custom flex PCB for the CCD biasing and clocking and a coaxial cable for the video signals (Figure ~\ref{fig2}).

The CCD control electronics is a standard ESO CCD controller (NGC), operating with the ESO control software, coupled with custom electronics and optimized for SOXS spectroscopic operations.
Several readout modes and speeds were defined and tested to optimize the SOXS operations for different scientific purposes.

\section{The TNG facilities}

The laboratory facility is based on an optical bench equipped with a Xenon lamp, filter wheels to select the wavelength, an integrating sphere and a calibrated diode to measure the flux.
An ad-hoc mechanical adapter allows for the coupling of the UV-VIV SOXS camera with the TNG facilities in order to perform all the required tests (Figure ~\ref{fig5}).

\begin{figure} [H]
   \begin{center}
   \begin{tabular}{c} 
   \includegraphics[height=9 cm]{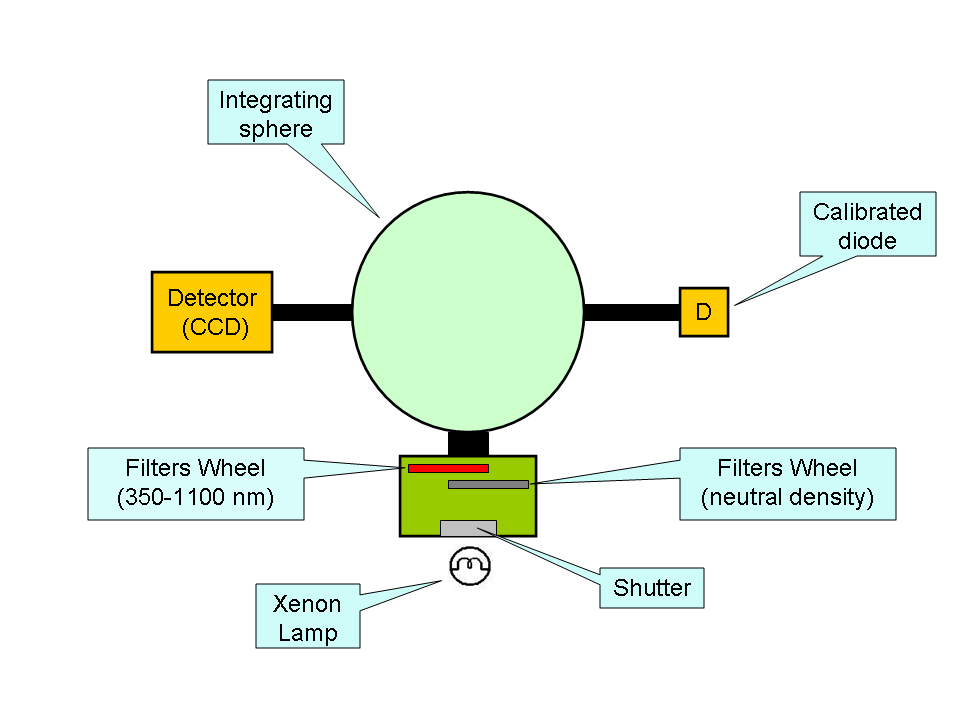}
   \end{tabular}
   \end{center}
   \caption[VIS Camera] 
   { \label{fig3} The scheme of  the laboratory facilities.}
   \end{figure}

The first test with the scientific CCD was done with the VIS chamber coupled with the TNG characterization facilities, in order to tune up the CCD parameters and to calculate the gain and the readout noise. In these test, we measured the gain and readout noise at different readout modes and readout speeds.
The optical bench for the detector tests is made up by some electronics apparatus, located in the optical laboratory, and the control computers, located in the annexed control room. 
The source is a Xenon lamp and the light goes through the shutter, the filter wheels and the integrating sphere and come to the SOXS UV-VIS detector (Figure ~\ref{fig3}).
The filter wheels make available the wavelength from 350 to 1100 nm and the possibility to reduce the flux with neutral filter from ND 0.1 to 4.


   \begin{figure} [H]
   \begin{center}
   \begin{tabular}{c} 
   \includegraphics[height=9 cm]{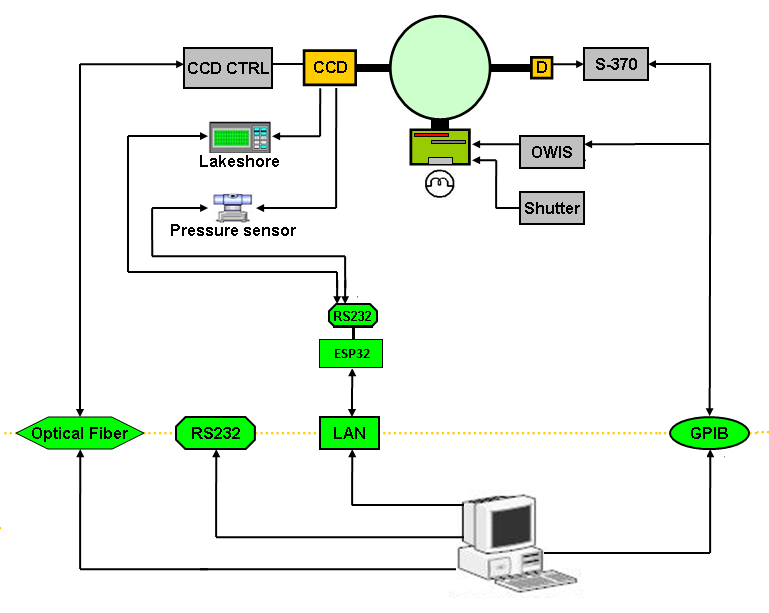}
   \end{tabular}
   \end{center}
   \caption[VIS Camera] 
   { \label{fig4} Setup and test environment (architecture).}
   \end{figure}

   \begin{figure} [H]
   \begin{center}
   \begin{tabular}{lcc} 
   \includegraphics[height=10 cm]{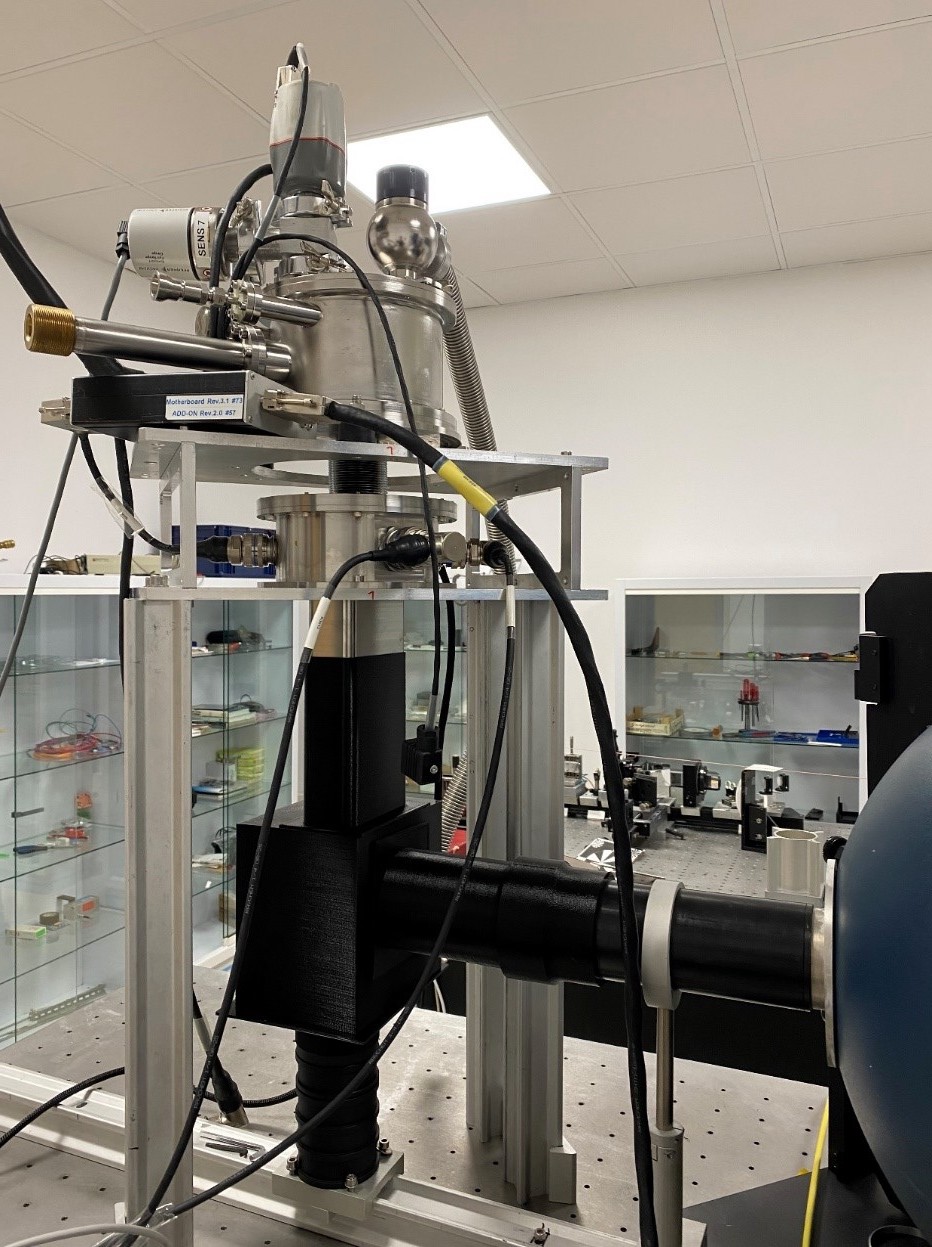}
   \includegraphics[height=10 cm]{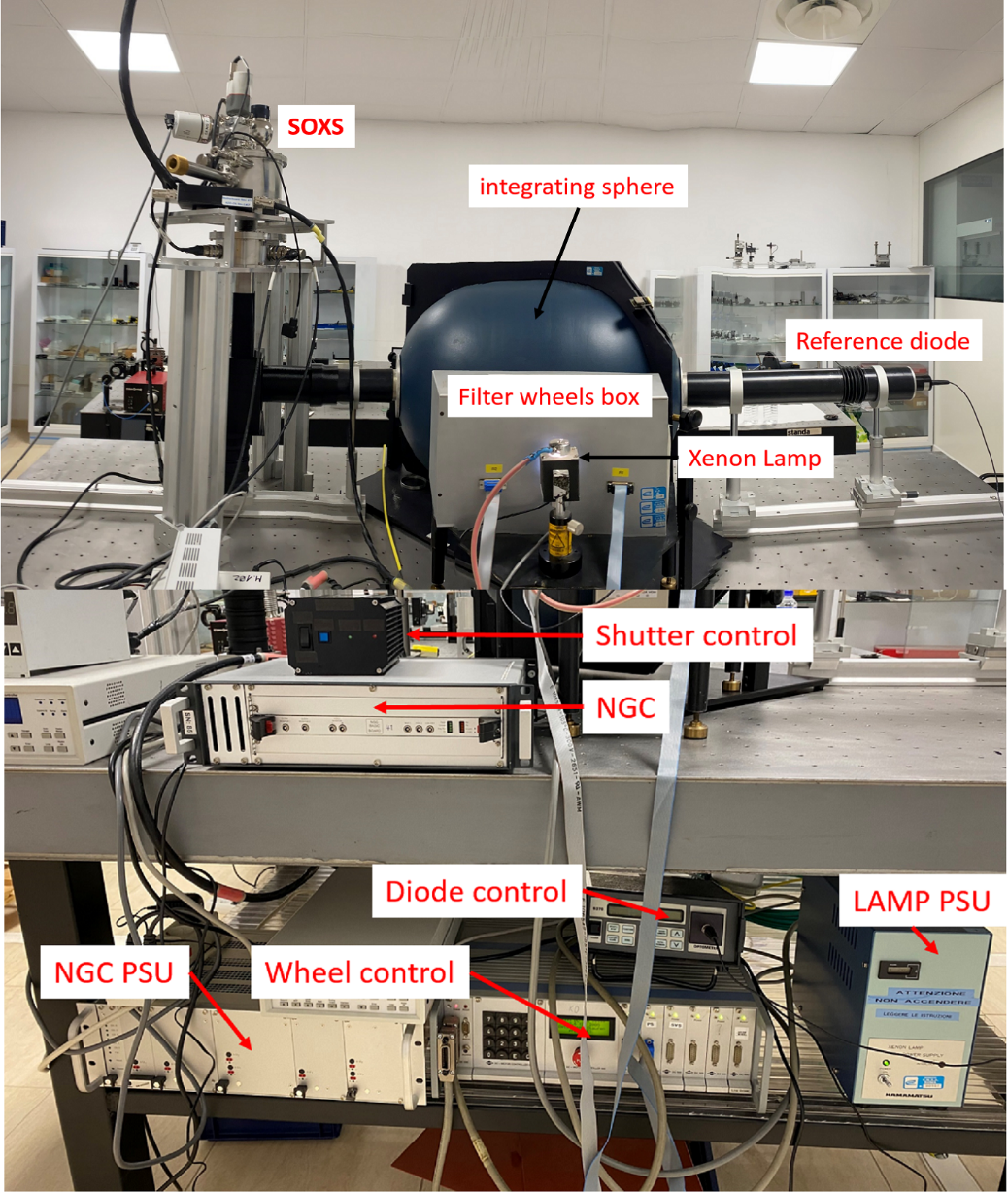}
   \end{tabular}
   \end{center}
   \caption[VIS Camera] 
   { \label{fig5} The coupling of the chamber (left) and The test system (right).}
   \end{figure}

\section{The test results}

The vacuum and cryogenic tests on the chamber show a good operation: the degassing rate is negligible and the LN2 consumption is as expected.
In these tests, we tested the scientific CCD in the same cryogenic environment foreseen for SOXS.  

\subsection{Test environment} 
Configuration:
\begin{itemize}
     \item [-]NGC controller with video-bias and Clock cables
    \item [-]Preamplifier 
    \item [-]SOXS camera and scientific CCD
    \item [-]CFC cooling system with the TeePee controller
    \item [-]Lakeshore temperature controller
    \item [-]Optical laboratory facilities
\end{itemize}

Tests:
\begin{itemize}
     \item [-]Bias acquisition and video offset settings
    \item [-]Flat field acquisition at different signal level
    \item [-]Flat field at different wavelength (350-1100 nm)
    \item [-]Gain measurements
    \item [-]Readout noise measurement
    \item [-]Linearity measurements
    \item [-]Test of different readout mode
    \item [-]Test of different binning (1X2, 2X2, 1X4, 2X4)
\end{itemize}

\subsection{Gain Measurements}  
For the gain measurements (conversion factor, CF) we used the method based on the assumption that the CCD have a fixed electron noise and the photons measured by the CCD follow the Poisson statistic. The CCD is uniformly illuminated at different levels of light and we take several bias and pairs of flat fields for each exposure time. The exposure time is increased and the signal level is measured in ADUs until saturation is reached. 

\subsection{Readout noise}  
The readout noise for the different speeds and gains was calculated on the bias images. These measurements will be repeated when the acquisition system will be assembled with the spectrograph in the dummy derotator.

\subsection{Linearity}  
\begin{figure} [H]
   \begin{center}
   \begin{tabular}{c} 
   \includegraphics[height=10 cm]{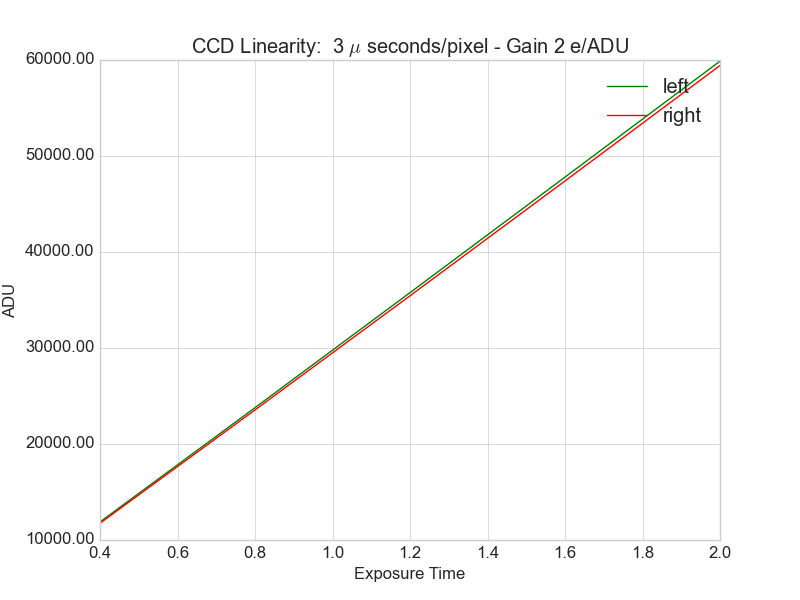}
   \end{tabular}
   \end{center}
   \caption[VIS Camera] 
   { \label{fig6} CCD linearity: speed=3 $\mu$sec/pixel - Gain 2e/ADU}
   \end{figure}

\begin{figure} [H]
   \begin{center}
   \begin{tabular}{lcc} 
   \includegraphics[height=6.5 cm]{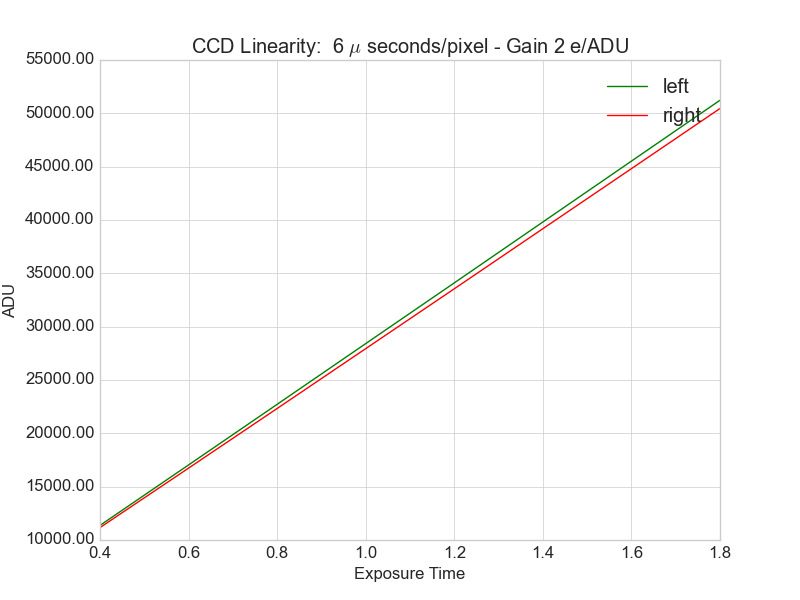}
    \includegraphics[height=6.5 cm]{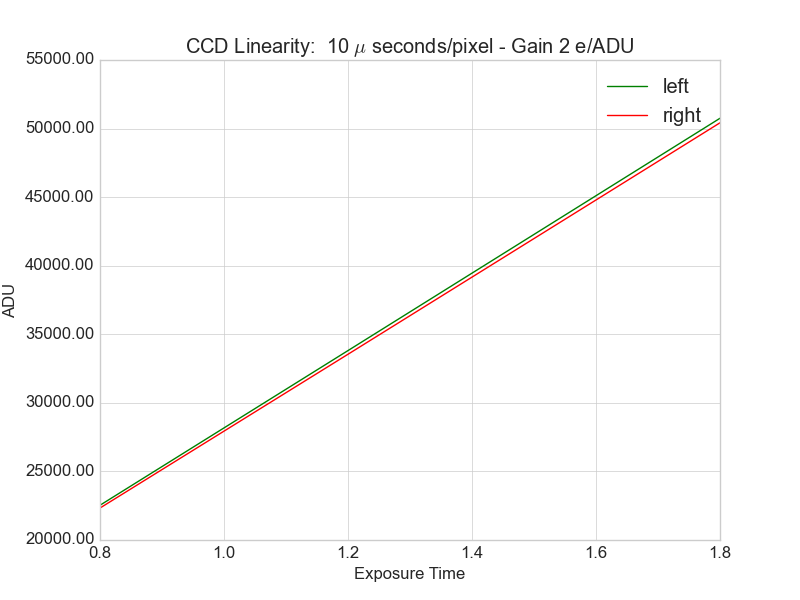}
   \end{tabular}
   \end{center}
   \caption[VIS Camera] 
   { \label{fig7} CCD linearity: speed=6$\mu$sec/pixel (left) and 10$\mu$ sec/pixel(right)  sec/pixel - Gain 2e/ADU.}
   \end{figure}

The linearity of the CCD vs. the incoming signal was calculated for the different speeds and gains and the results are shown in Figure ~\ref{fig6} and in Figure ~\ref{fig7}. The two (red and green) lines of the figures, shown the linearity of the two outputs of the CCD.

\subsection{Binning}  
Several binnings was tested (1X2, 2X2, 1X4, 2X4) and the results was as expected (no changes in readout noise and gain).

\subsection{Windowing of the CCD}  
\begin{figure} [H]
   \begin{center}
   \begin{tabular}{c} 
   \includegraphics[height=8.5 cm]{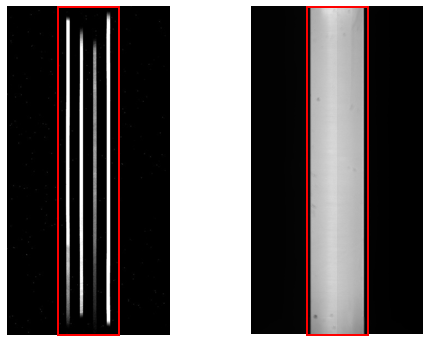}
   \end{tabular}
   \end{center}
   \caption[VIS Camera] 
   { \label{fig8} On the left is shown the theoretical image on the CCD, on the right the flat field obtained in laboratory.}
   \end{figure}

The optical design foresee that the illuminated area is the central part of the active area of the CCD (740X4096 pixels). This means that the readout speed can be reduced by using the windowing of the CCD with a notable increasing of the performances of the acquisition system. In Figure ~\ref{fig8} are shown the simulated image of the spectra in the CCD on the left and the flat field obtained with the CCD chamber with the test bench used to test the detector system.  

   The measurement of the CCD performances shows a good linearity, low readout noise and allow for different readout modes with various gain and readout speed, in order to can select the modes that best fit with the observation modes.
   
   \section{INTEGRATION WITH UV-VIS}
   The chamber was coupled with the UV-VIS arm of SOXS spectrograph at the end of June 2022, the first tests were done and the data will be analyzed in the next weeks. In the figure below are shown a picture of the UV-VIS spectrograph and one of the spectra taken during the test campaign.
   
   \begin{figure} [H]
   \begin{center}
   \begin{tabular}{c} 
   \includegraphics[height=10 cm]{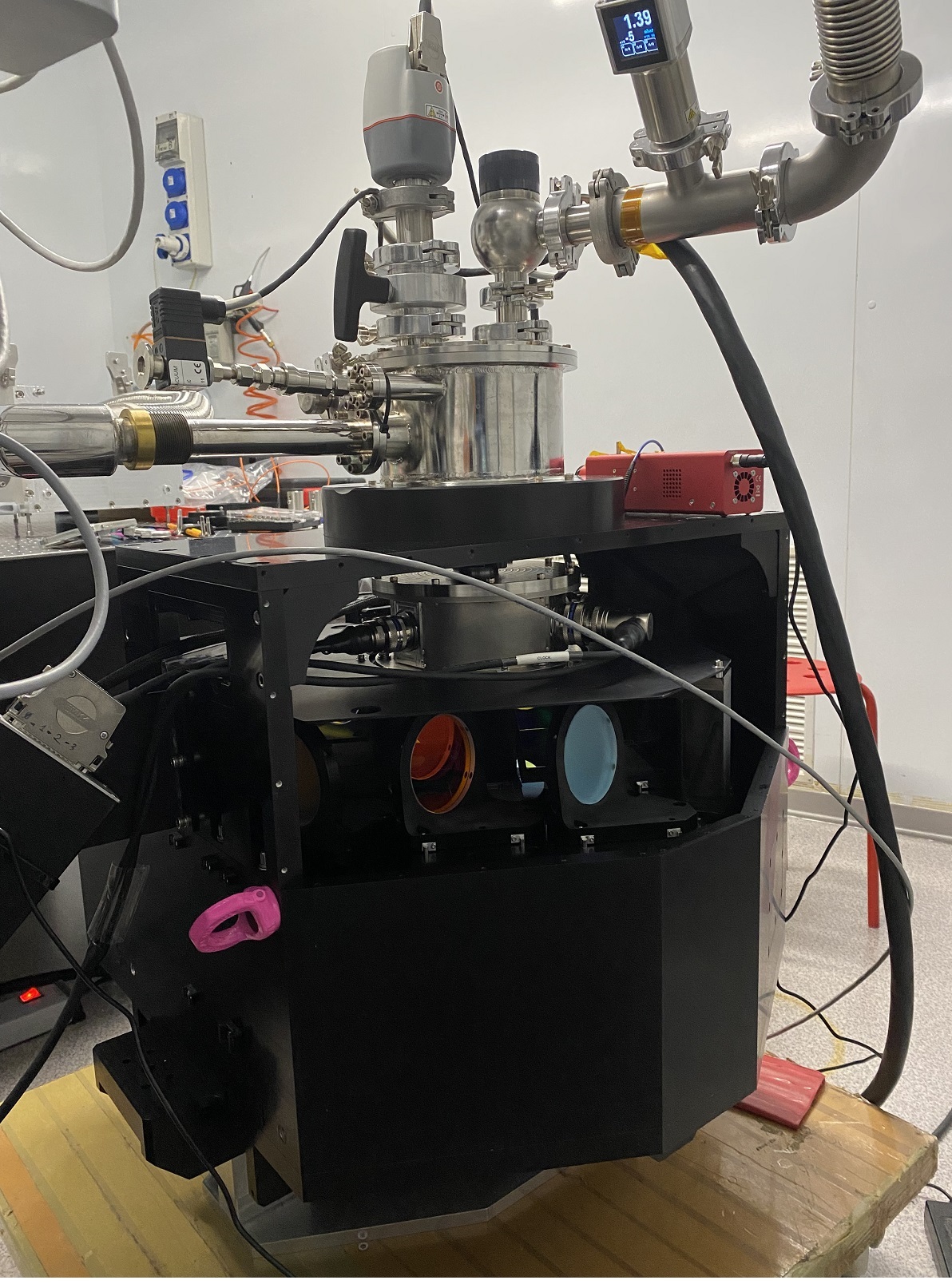}
   \includegraphics[height=10 cm]{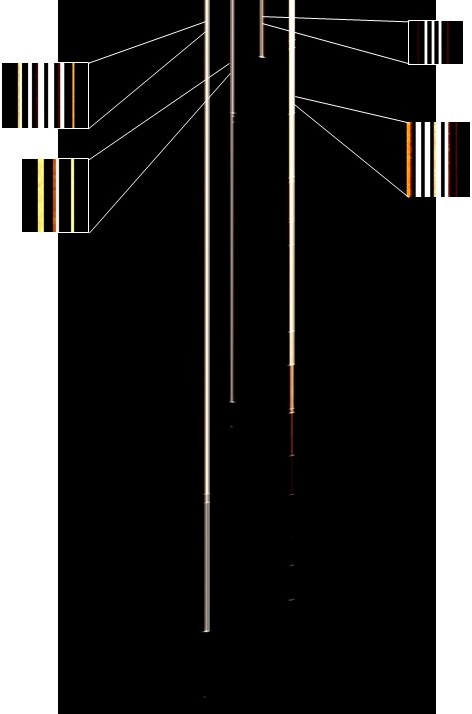}
   \end{tabular}
   \end{center}
   \caption[VIS Camera] 
   { \label{fig10} The UV-VIS arm and an example of one of the UV-VIS spectra.}
   \end{figure}
   
\subsection{Conclusions}
The tests on the final version of VIS detector system with the scientific CCD show that the acquisition system works as expected. The test in the characterization laboratory was done to optimize the performances of the system in term of readout noise and data throughput, and provide the measurement of gain and readout noise.
The final test on the VIS detector system coupled with the spectrograph was done in June 2022 and the results will be analyzed in the next weeks to characterize the entire UV-VIS system.

\acknowledgments 
 
A special acknowledgement to the European Southern Observatory for the support provided and for the availability to share its knowledge and to allow for the use of the ESO laboratories in Garching. 


 


\bibliography{VIS} 
\bibliographystyle{spiebib} 
\end{document}